\documentclass{aa}

\usepackage{graphicx}
\usepackage{color}
\usepackage{ulem}

\newcommand\fonction{EACF}
\def\muHz{\,$\mu$Hz}
\newcommand\temps{t}
\newcommand\sampling{\delta t}
\newcommand\res{\delta\nu}
\newcommand\dnu{\Delta\nu}
\newcommand\deltanunu{\Delta\nu(\nu)}
\newcommand\dnumoy{\langle\Delta\nu\rangle}
\newcommand\numax{\nu\ind{max}}
\newcommand\dnuc{\dnu\ind{c}}
\newcommand\narrow{0.75}
\newcommand\identideg{\mathcal{T}_{01}}
\newcommand\deltadeg{\delta_{01}}
\newcommand\Deltan{\Delta n}
\newcommand\HWHM{\delta\nu\ind{env}}
\newcommand\larg{\delta\nu\ind{H}}

\newcommand\largpic{\delta\tau}
\newcommand\centre{\nu\ind{c}}
\newcommand\degre{\ell}
\newcommand\nombre{N\ind{H}}
\newcommand\ndt{N\ind{t}}

\newcommand\HBR{\mathcal{R}}
\newcommand\niveau{\sigma}
\newcommand\nvaleur{\mathcal{N}}
\newcommand\level{\mathcal{P}}
\newcommand\redui{\mathcal{X}}
\newcommand\amp{\mathcal{A}}
\newcommand\ampdnu{\mathcal{A}_{\Delta\nu}}
\newcommand\ampdt{\mathcal{A}_{\delta t}}
\newcommand\ampmax{\mathcal{A}\ind{max}}
\newcommand\ampauto{\mathcal{A}\ind{auto}}
\newcommand\ampfit{50}
\newcommand\ampdeg{200}
\newcommand\amppetit{300}
\newcommand\bonintervalle{\Delta\tau}
\newcommand\Tdnu{{\tau_{\Delta\nu}}}
\newcommand\Tdnuc{\Tdnu\ind{c}}
\newcommand\coefauto{G}
\newcommand\correl{\mathcal{C}}
\newcommand\correlc{\mathcal{C_{\pm}}}
\newcommand\hilbert{\tilde\mathcal{H}}
\newcommand\filtre{\mathcal{F}}
\newcommand{\ind}[1]{_{\mathrm{#1}}}

\newcommand\sinc{\mathrm{sinc}}
\newcommand\diff{\, \mathrm{d}}

\newcommand\Kepler{Kepler}
\begin{document}
\title{
On detecting the large separation in the autocorrelation of stellar oscillation times series~\thanks{The CoRoT space mission, launched on 2006 December 27, was developed and is operated by the CNES, with participation of the Science Programs of ESA, ESA's RSSD, Austria, Belgium, Brazil, Germany and Spain.}}
\titlerunning{Large separation and autocorrelation function}
\author{%
B. Mosser\inst{1}\and
T. Appourchaux\inst{2}}

\institute{LESIA, CNRS, Universit\'e Pierre et Marie Curie, Universit\'e Denis Diderot, Observatoire de Paris, 92195 Meudon cedex, France\\
\email{benoit.mosser@obspm.fr}
\and
Institut d'Astrophysique Spatiale, UMR8617, Universit\'e Paris XI, B\^atiment 121, 91405 Orsay Cedex, France
}
\date{Received 22 July 2009 / Accepted 7 October 2009}

\abstract
{The  observations carried out by the space missions CoRoT and Kepler provide a large set of asteroseismic data. Their analysis requires an efficient procedure first to determine if a star reliably shows solar-like oscillations, second to measure the so-called large separation, third to estimate the asteroseismic information that can be retrieved from the Fourier spectrum.}
{In this paper we develop a procedure based on the autocorrelation of the seismic Fourier spectrum that is capable of providing measurements of the large and small frequency separations. The performance of the autocorrelation method needs to be assessed and quantified. We therefore  searched for criteria able to predict the output that one can expect from the analysis by autocorrelation of a seismic time series. }%
{First, the autocorrelation is properly scaled to take into account the contribution of white noise. Then we use the null hypothesis H$_{0}$ test to assess the reliability of the autocorrelation analysis. Calculations based on solar and CoRoT time series are performed to quantify the performance as a function of the amplitude of the autocorrelation signal.}%
{We obtain an empirical relation for the performance of the autocorrelation method. We show that the precision of the method increases with the observation length, and with the mean seismic amplitude-to-background ratio of the pressure modes to the power 1.5$\pm$0.05. We propose an automated determination of the large separation, whose reliability is quantified by the H$_{0}$ test. We apply this method to analyze red giants observed by CoRoT.  We estimate the expected performance for photometric time series of the Kepler mission. We demonstrate that the method makes it possible to distinguish $\degre=0$ from $\degre=1$ modes.}%
{The envelope autocorrelation function (\fonction) has proven to be very powerful for the determination of the large separation in noisy asteroseismic data, since it enables us to quantify the precision of the performance of different measurements: mean large separation, variation of the large separation with frequency, small separation and degree identification.}

\maketitle
\voffset = 1.5cm
\section{Introduction\label{introduction}}

Asteroseismology is known to be an efficient tool to analyze the stellar interior and to derive the physical laws that govern stellar structure and evolution. It benefits nowadays from high-performance photometric data provided by the space missions CoRoT (\cite{2006ESASP1306...33B}) and Kepler (\cite{2007CoAst.150..350C}). The amount of data is much higher than from the earlier ground-based observations, even with the recent multi-site ground-based observations (\cite{2008ApJ...687.1180A}), since space-borne instruments are able to simultaneously record long time series on numerous targets. The data analysis then must be efficient enough to rapidly  extract seismic information from hundreds to thousands of stars.

This task is principally carried out on the frequency pattern of the eigenmodes propagating inside the stars.
For targets showing solar-like oscillations, this pattern follows the asymptotic relation of \cite{1980ApJS...43..469T} providing eigenfrequencies nearly equally spaced by $\dnu/2$.
The eigenfrequency of radial order $n$ and degree $\degre$ expresses
$\nu_{n,\degre} \ \simeq \ \left[ n + {\degre / 2} + \varepsilon \right] \dnu - \degre (\degre +1) D_0$,
$\dnu$ being called the large separation, $D_0$ giving a measure of the small separation,
and $\varepsilon$ a constant term.
The determination of the large separation $\dnu$ is the first step of any seismic analysis.
If the signal-to-noise ratio is high enough, $\dnu$ can be detected by eye in the power spectrum.
In many cases, this is not possible, and the determination of $\dnu$ requires sophisticated tools, as was the case for the first correct determination of the large separation of Procyon (\cite{1998A&A...340..457M}) and of the first CoRoT target observed with Doppler measurements (\cite{2005A&A...431L..13M}).
For observations dealing with a single target, the tools used for the determination of $\dnu$ are usually unautomated and involve parameters specific to the target. Most often, they require the visual inspection of an image or a graph obtained by transforming the Fourier spectrum using the asymptotic relation cited above (\'echelle diagram or comb response). This step can be automated, but with great care, since the higher order terms of \cite{1980ApJS...43..469T} complicate the stacking, as does, for example, the variation of the large separation reported in many asteroseismic targets (e.g. \cite{2008A&A...488..635M}).

With the advent of space photometric missions, the use of pipelines for the automatic detection of $\dnu$ is becoming mandatory (\cite{2009arXiv0907.1139M}) since  many  targets have a low signal-to-noise ratio. A test to determine if the large separation is reliably detected is highly desirable, and a way to estimate the asteroseismic content of a high-precision photometric time series will be very helpful.

This paper proposes an original way to address these issues. It is based on a first report  by \cite{2006MNRAS.369.1491R} (hereafter RV06), who analyse solar-like oscillations via the square of the autocorrelation of the time series, calculated as the Fourier spectrum of the filtered Fourier spectrum. RV06 state that the method is useful when faced with low signal-to-noise ratio data, and might be useful in obtaining information about a star even when individual frequencies cannot be extracted.
\cite{roxburgh2009} (hereafter R09) shows that it is possible, with basic and rapid computations, to attain more complex objectives, such as measurement of the variations of the large separation with frequency.

Since it provides a rapid measurement of the large separation, the autocorrelation method fits perfectly with the main asteroseismic objective of the \Kepler\ mission, the large separation being  used as an independent measurement in extracting the radius of stars hosting exoplanets, as in \cite{2009ApJ...700.1589S}. The autocorrelation achieves this goal without fitting a complex mode pattern to the stellar power spectrum.  Therefore, it provides a simple tool to estimate the asteroseismic information of a Fourier spectrum or to use with \Kepler, which will produce numerous time series of stellar targets.

We propose to quantify the relevance of the autocorrelation method with the null hypothesis, and to determine simple criteria to assess its efficiency and predictive power when analyzing an oscillation spectrum with a low signal-to-noise ratio. The method is also useful for extrapolating the performance obtained with a short time series to that obtained with a 4-year long time series, as will be provided by the \Kepler\ mission.
The analysis relies on photometric time series as observed by CoRoT (\cite{2006ESASP1306...33B}), plus simulations based on these CoRoT spectra with the addition of noise. It also includes simulations derived from a solar oscillation spectrum observed in photometry by the VIRGO/SPM instrument of the SOHO mission.

Section~\ref{autocorrelation} introduces the envelope autocorrelation function (\fonction) and the way we scale it to properly account for the noise contribution. We show in Section~\ref{analysis} how the value of the main autocorrelation peak varies with different global parameters of the stellar oscillation spectrum. A crucial parameter is the mean seismic height-to-background ratio $\HBR$, representing the smoothed height of the seismic power spectral density compared to the background.
We introduce in Section~\ref{performance} the H$_{0}$ test, that allows us to examine and to quantify the performance of the method. The value of the \fonction\ gives a reliable criterion to estimate the seismic output, from the determination of the mean large separation when the signal is poor to the possibility of precise mode fitting in other cases. Discussion of various cases is presented in Section~\ref{discussion}. We propose an automated determination of the large separation; using the H$_{0}$ test, we can quantify the reliability of this method. Section~\ref{conclusion} is devoted to conclusions.

\section{Autocorrelation\label{autocorrelation}}

\subsection{Calculation}

RV06 proposes to perform the autocorrelation of the seismic time series as the Fourier spectrum of the filtered Fourier transform of the time series. This directly gives the amplitude of the envelope of the autocorrelation function, as shown in the Appendix.
Instead of the canonical form,
\begin{equation}\label{correl}
\correl (\tau) = \int\! x(t) \, x(t+\tau) \diff t = \int\! X(\nu) X^* (\nu) {\rm e}^{{\rm i} 2\pi\nu \tau} \diff \nu
\end{equation}
with $X(\nu)$ the Fourier transform of $x(t)$, the autocorrelation with a  filter $\filtre$ of width $\larg$ centered on $\centre$ can be written:
\begin{equation}\label{correlcos}
\correl
= \int_{\centre-\larg}^{\centre+\larg}
\!\!\!\!\!\!
X(\nu) X^* (\nu)
\ \filtre (\nu) \
{\rm e}^{{\rm i} 2\pi\nu \tau} \diff \nu .
\end{equation}
We deal with the dimensionless square module of the autocorrelation:
\begin{equation}\label{autocorrel}
{\amp}^\star = \left| \correl (\tau)^2\right|  /  \left| \correl (0)^2\right|.
\end{equation}
The choice of square module has no impact on the results presented below, but proved to be more convenient in many cases, such as the observed linear increase of ${\amp}^\star$ with the observing time (see Eq.~(\ref{prop})).

\subsection{Noise scaling}

In order to compare different cases, it is preferable to express the amplitude of the autocorrelation signal in noise units. The mean noise level in the autocorrelation can be derived from the fact that the noise statistic is a $\chi^2$ with 2 degrees of freedom. It is expressed in the general case as:
\begin{equation}\label{ns_un}
  \niveau = {2\over \ndt} \ {\langle\filtre^2\rangle \over \langle\filtre\rangle^2}
\end{equation}
with  $\ndt$ the number of points in the time series.
The noise level $\niveau$ is inversely proportional to the number $\nombre$ of frequency bins selected in the filter, when $\nombre$ is measured in a Fourier spectrum at the exact frequency resolution $\res = 1/T$, $T$ being the length of the observation, without oversampling. For a cosine filter (or Hanning function) of full-width at half-maximum $\larg$ centered on $\centre$, one gets:
\begin{equation}\label{noiselevelN}
   \niveau\ind{H} = {3\over 2\, \nombre}.
\end{equation}
With such a cosine filter and the resulting noise level, we define the \fonction :
\begin{equation}\label{normalisation}
  \amp = {\amp}^\star / \niveau\ind{H}.
\end{equation}
We note  that $\ampdnu = \amp (\Tdnu) $, the amplitude of the first peak in the autocorrelation function, at a time shift $\Tdnu = 2/\dnu$. The first peak of the autocorrelation (Fig.~\ref{correl_env}) is the signature of the autocorrelation of a seismic wavepacket after crossing the stellar diameter twice. As shon in RV06, measuring the time delay $\Tdnu$ of this peak allows us to measure the large separation.

\begin{table*}
\caption{$\HBR$  and parameters of the p mode envelope for CoRoT targets}\label{tablehbr}
\begin{tabular}{rllrllccccc}
\hline
star&type&$m\ind{V}$& t & $\numax$ & $\HWHM$ & $\dnumoy$ & $\alpha$ & $\gamma$ & $\HBR$ & $\ampmax$\\
   &    &           & (day) &\multicolumn{2}{c}{\dotfill (mHz) \dotfill}& (\muHz) & \\
\hline
HD49385$^a$   & G0IV& 7.41 & 136.9 &   1.00 & 0.54 & 56 & 0.95 & 9.7& 0.90 &  452\\
HD49933$^{b1}$& F5V & 5.78 & 136.9 &   1.79 & 0.86 & 85 & 1.10 &10.1& 0.75 &  562\\
HD49933$^{b2}$&     &      & 60.7  &        &      &    &      &    &      &  237\\
HD175726$^c$  & G0V & 6.72 & 27.2  &   2.05 & 0.82 & 97 & 1.20 & 8.5& 0.11 &  6.9\\
HD181420$^d$  & F2V & 6.57 & 156.6 &   1.60 & 0.76 & 76 & 1.15 &10.3& 0.42 &  242\\
HD181906$^e$  & F8V & 7.6  & 156.6 &   1.92 & 0.88 & 85 & 1.00 &10.3& 0.15 &   47\\
HD181907$^f$  &G8III& 5.8  & 156.6 & 0.0286 &0.0176& 3.5& 1.10 & 5.0& 1.99 &   40\\
Sun/VIRGO$^g$& G2V &      & 182.1 &   3.25  & 1.04 & 135& 1.30 & 8.0& 2.50 & 7.5\,10$^3$\\
\hline
\end{tabular}

$\numax$ is the location of the frequency of maximum power; $\HWHM$ is the full-width at half-maximum of the mode envelope; $\dnumoy$ is the mean value of the large separation; $\alpha$ represents the optimized filter width, in unit $\HWHM$;  $\gamma$ represents $\HWHM$ in unit $\dnu$; $\HBR$ measures the mean seismic amplitude in the time series compared to the noise, by the ratio in the Fourier spectrum, at the maximum-oscillation frequency, of the smoothed mode height to the background power density;  $\ampmax$ measures the \fonction.

References:
$^a$\cite{deheuvels2009};
$^{b1}$\cite{benomar2009};
$^{b2}$\cite{2008A&A...488..705A};
$^c$\cite{mosser2009};
$^d$\cite{barban2009};
$^e$\cite{garcia2009};
$^f$\cite{carrier2009};
$^g$\cite{1997SoPh..170....1F}.

\end{table*}

\begin{figure}
\centering
\includegraphics[width=8.5cm]{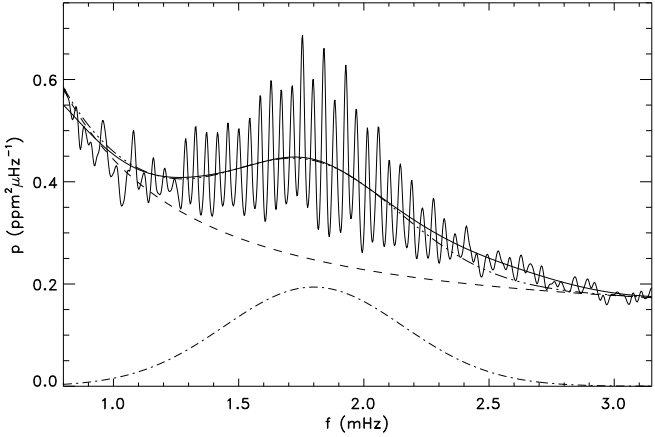}
\caption{Contributions to the smoothed power density distribution for HD49933. The oscillation spectrum was slightly and severely smoothed (solid thin and thick lines). The dashed line represents the contributions of granulation and photon noise. The dash-dot lines account for the Gaussian modeling of the seismic envelope and the total contribution.
\label{profilsbr}}
\end{figure}

\section{Analysis\label{analysis}}

We tested the variation of $\ampdnu$ with various parameters, in order to determine the relevant ingredients contributing to this signal. We based the analysis on solar data obtained with the VIRGO/SPM instrument  onboard SOHO (\cite{1997SoPh..170....1F}), and on the CoRoT data provided on the solar-like targets HD49933 (Appourchaux et al 2008), HD49385 (\cite{deheuvels2009}), HD175726 (\cite{mosser2009}), HD181420 (\cite{barban2009}) and HD181906 (\cite{garcia2009}). We also include the red giant HD181907 observed by CoRoT (\cite{carrier2009}). All these targets are presented in Table~\ref{tablehbr}. We also considered a set of red giants observed in the exoplanetary field of CoRoT, already analyzed by \cite{hekker2009}.

\subsection{Seismic amplitude-to-background ratio}

The strength of the autocorrelation of the time series depends on the ratio of the mean seismic amplitude compared to all other signal and noise. We can derive this signal-to-noise ratio in the time series from the ratio estimated in the oscillation spectrum. This ratio in the Fourier spectrum does not depend on the frequency resolution when the modes are resolved, i.e. the observation time is longer than the mode lifetime. In order to remove the influence of unknown parameters, such as the star inclination or the mode lifetime, we have to consider the ratio $\HBR$ of the mode height to the background power density, at the maximum-oscillation frequency, in a smoothed power density spectrum  (Fig.~\ref{profilsbr}).

In order to estimate the background power, we have modeled the Fourier spectra with three components as in  \cite{2008Sci...322..558M}: a low-frequency Lorentzian-like profile, a Gaussian mode envelope and a high-frequency noise. Figure~\ref{profilsbr} shows this modeling for HD49933.
The smoothed power density depends on the filter width. In order to avoid Gibbs phenomenon-like structures, a Gaussian filter has to be preferred to a boxcar average. The width has to be proportional to the large separation: a value of $3\,\dnu$ provides the optimum smoothing and limits the influence of the varying background level. Since, at this stage, the large separation is a priori unknown, the value of the filter width can be estimated with the help of the relation found between the large separation and the maximum-power frequency derived from the solar-like CoRoT targets:
\begin{equation}\label{scalingnumax}
\dnu \simeq (0.24 \pm 0.05)\  \numax^{0.78\pm 0.045} \ \ \ \hbox{(frequencies in $\mu$Hz)}.
\end{equation}
Table~\ref{tablehbr} gives $\HBR$ calculated for a set of CoRoT targets with solar-like oscillations, with the location $\numax$ of the maximum PSD and the full-width at half-maximum $\HWHM$ of the pressure mode envelope. The precision of the determination of $\HBR$ derived from these CoRoT targets is about 15\%.

\begin{figure}
\centering
\includegraphics[width=8.5cm]{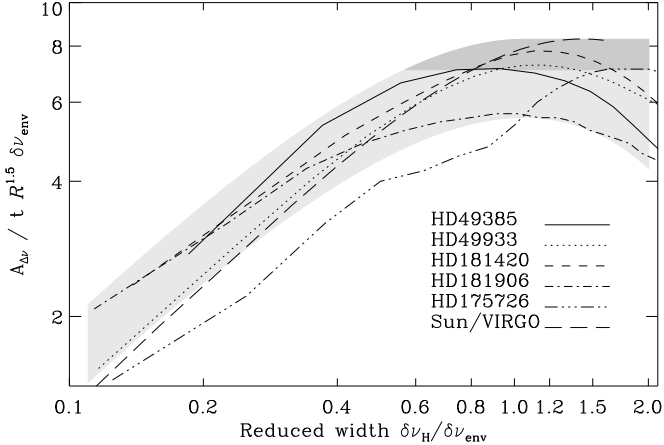}
\caption{Variation of the reduced amplitude $\ampdnu\,t^{-1} \,\HBR^{-1.5}\, \HWHM^{-1}$ as a function of the reduced filter width $\larg / \HWHM$. The light grey region indicates the validity of the global mean fit given by Eq.~(\ref{prop}), within a $\pm$20\,\% precision. The dark grey region indicates the location of the maxima, within a $\pm$15\,\% precision except for the double star HD 181906.
\label{estimperflarg}}
\end{figure}

\subsection{\fonction\ as a function of time, filter width and signal-to-noise ratio}

The scaling (Eq.~\ref{normalisation}) permitted us to perform different treatments in order to analyze how the \fonction\ varies with the observing time $t$, the filter width $\larg$, the full-width at half-maximum of the mode envelope $\HWHM$ and $\HBR$.
With a linear dependence of $t$ and the introduction of the reduced width $\redui = \larg/ \HWHM$, we found:
\begin{equation}\label{prop}
   \ampdnu
   \simeq  19.2\
\HBR^{1.5}
\left[{ \temps \over 1\, \mathrm{day}}\right]\,
\left[{ \HWHM\over 1\, \mathrm{mHz}}\right]\
\redui\,  \exp\left(-{\redui\over 1.05}\right).
\end{equation}

The amplitude $\ampdnu$ increases linearly with time, since $\niveau$ is inversely proportional to the observation time $\temps$ according to Eq.~(\ref{noiselevelN}). As an important consequence, despite the limited lifetime of the modes, the precision of the seismic autocorrelation diagnosis increases linearly with $\temps$. This increase corresponds in fact to a decrease of $\sigma$ and therefore cannot saturate. The factor 1.05 in Eq.~(\ref{prop}) is derived from the comparison between a Gaussian envelope and the Hanning filter: the best fit with such a filter requires a full width at half maximum equal to 1.05 times the one of the mode envelope.

Figure~\ref{estimperflarg} shows the global fit, valid for photometric data of solar-like stars obtained with CoRoT or with VIRGO/SPM onboard SOHO. All values are fit within $\pm 20$\%,  when $\larg\le 2\, \HWHM$, except the amplitudes for HD175726, which is the target with the lowest $\HBR$; however, the maximum amplitude for this star agrees with the others.

We have verified that the exponent of the $\HBR$ dependence that minimizes the dispersion of the different curves in Fig.~\ref{estimperflarg} is $1.5\pm 0.05$. A theoretical analysis should be performed to assess this result. Such a work requires one to take into account the link between $\HBR$ and the star inclination, the mode lifetime and the stellar noise.

\subsection{Maximum autocorrelation signal}

From Eq.~(\ref{prop}), we can derive the maximum autocorrelation signal, obtained for $\larg = \alpha \, \HWHM$. It varies as:
\begin{equation}\label{propmax}
\ampmax \simeq  7.0 \ \alpha \ \HBR^{1.5}
\left[{ \temps \over 1\, \mathrm{day}}\right]\
\left[{ \HWHM\over 1\, \mathrm{mHz}}\right].
\end{equation}
The parameter $\alpha$ is derived from the location of the maximum signal (Fig.~\ref{estimperflarg}). If $\alpha$ has not be determined, it should be replaced by its typical value $\alpha\simeq 1.05$ as used in Eq.~(\ref{prop}).

Figure~\ref{estimperflarg} helps to identify $\ampmax$. For all solar-like single stars but HD181906, the agreement with Eq.~(\ref{propmax}) is better than $\pm$15\%.
The fact that HD181906 shows the lowest maximum among solar-like stars is certainly due to its binarity (\cite{bruntt2009}): $\HBR$ and $\ampmax$ are corrupted by the unknown contribution of the companion. The observed value of $\ampmax$ for the red giant HD181907, not shown, is 2 times lower than expected. This is clearly related to the narrow envelope of its oscillation spectrum, expressed by $\gamma = 5$ compared to a mean value of 10 for solar-like stars (Table~\ref{tablehbr}). The number of observed p modes is then twice as small and the \fonction\ is reduced.

\section{Performance\label{performance}}

The scaling of $\ampmax$ with Eq.~(\ref{normalisation}) allows us to test the reliability of the detection of the large separation with the H$_{0}$ test, and then to estimate the scientific output of the \fonction. The null hypothesis, term first coined by the geneticist and statistician Ronald Fisher in \cite{fisher1935}, consists here of assuming that the correlation is generated by pure white noise. If the \fonction\ is high enough, the H$_{0}$ hypothesis is rejected, implying that a signal might have been detected  (\cite{2004A&A...428.1039A}).

\subsection{H$_{0}$ test\label{testhzero}}

Assessing the reliability of the measurement of the large separation as proposed by RV06 implies applying a statistical test as the null hypothesis H$_{0}$. A priori information on the large separation may come from scaling laws (\cite{1983SoPh...82..469C}), or may be derived from the location of the maximum signal, or from the initial guess of the stellar fundamental parameters. The large separation is then searched  for over a range $\bonintervalle$. The number $\nvaleur$ of independent bins over the range $\bonintervalle$ depends on the width of the cosine filter. It is proportional  but not equal to the number of points $\nombre$ selected by the filter in the Fourier spectrum. It can be determined from the full width at half-maximum $\largpic$ of the autocorrelation peaks. Then, $\nvaleur$ is:
\begin{equation}\label{nombre_valeur}
    \nvaleur = {\bonintervalle \over \largpic}.
\end{equation}
Therefore the rejection of the H$_{0}$ hypothesis at probability level $\level$ implies a threshold  value:
\begin{equation}\label{threshold}
   \amp\ind{lim} \simeq - \ln (\level )+\ln\left({\bonintervalle \over \largpic}\right).
\end{equation}
This equation is only valid if $\level \ll 1$.  Equation~\ref{threshold} shows that the threshold increases with the searched range, but decreases with the resolution $\largpic$. We verified (see Eqs.~\ref{peakhanning} and \ref{demolargeur}) that $\largpic$ is related to the width $\larg$ of the cosine filter. Then, $\nvaleur$ is
\begin{equation}\label{nombrevaleur}
    \nvaleur = {1\over \beta} \ \bonintervalle \, \larg
\end{equation}
with $\beta= 0.763$.
This number $\nvaleur$ can be estimated, even if nothing is known about the target, since $\bonintervalle$ and $\larg$ are both function of the large separation. The interval $\bonintervalle$ where the autocorrelation peak is to be found is measured by the time shift $\Tdnu$. As a conservative value we may consider $\bonintervalle = \Tdnu = 2/ \dnu$.

As shown above, the width of the best filter giving the maximum autocorrelation signal is proportional to the width of the seismic mode envelope, $\larg = \alpha \,\HWHM$, and  the mode envelope also varies almost linearly with the the large separation, $\HWHM = \gamma\, \dnu$. This gives $\larg = \alpha \gamma \, \dnu$. As a consequence, independent of the large separation, the number of independent bins in the autocorrelation can be estimated by:
\begin{equation}\label{nombrevaleur2}
    \nvaleur = 2 \ {\alpha \gamma \over \beta}.
\end{equation}
Setting the mean optimum value of $\alpha$ to 1.05 and the mean value of $\gamma$ to 10, we obtain the number of independent bins to be considered in the \fonction, about 28. The threshold values for rejecting the H$_{0}$ hypothesis at level $\level=$ 1\% or 10\% are then respectively 8.0 and 5.7.

\begin{figure}
\centering
\includegraphics[width=8.5cm]{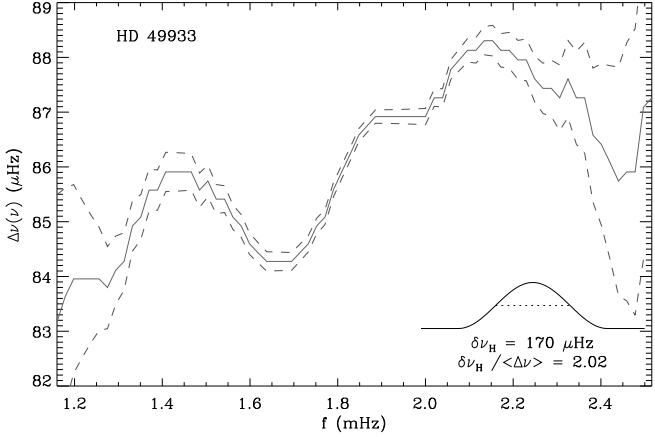}
\caption{Precision of the function $\deltanunu$ of HD49933. 1-$\sigma$ error bars are given by the dashed lines.
The inset shows the cosine filter, with a full-width at half-maximum equal to 2 times the mean large separation.
\label{autodeltanunu49933}}
\end{figure}
\begin{figure}
\centering
\includegraphics[width=8.5cm]{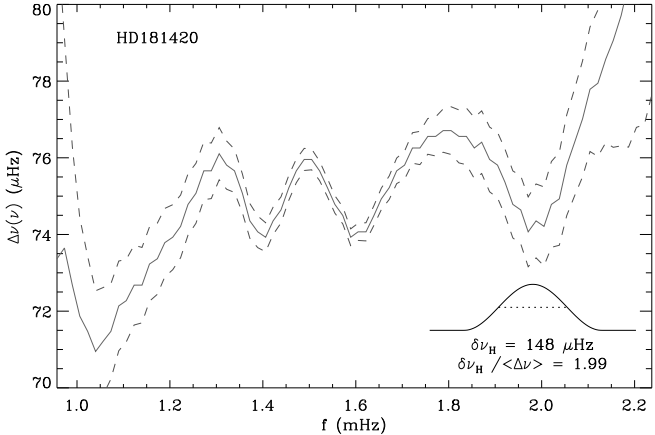}
\caption{Same as Fig.~\ref{autodeltanunu49933}, but for HD181420.
\label{autodeltanunu181420}}
\end{figure}

\subsection{Determination of the mean large separation $\dnumoy$}

The determination of the mean value of the large separation requires $\ampmax$ to be greater than a threshold value of about 8 for a detection at the 1\% rejection level. From  Eq.~(\ref{conditiondeux}), we then get an estimate of the relative precision of the mean large separation $\dnumoy$, integrated over a large frequency centered on the maximum-oscillation frequency:
\begin{equation}\label{conditionmoyenne}
   {\delta \dnumoy \over \dnumoy} \simeq   { 0.057\ \over  \ampmax}.
\end{equation}
At the detection limit $\ampmax = 8$, this gives a relative precision of about 0.6\,\%. The meaning of a mean value of the large separation is questionable, since most of the stars show significantly varying $\deltanunu$. These variations, integrated over the filter width, may limit the precision of $\dnumoy$, hence of an output value such as the stellar radius.

\subsection{Variation of the large separation with frequency}

With smaller values of $\larg$, it is possible to address the variation of the large separation with frequency as explained in R09.  Investigating in detail $\deltanunu$ requires a filter $\larg$ much narrower than  $\HWHM$, so that we can derive $\amp \simeq \ampmax \; e/\alpha \; \larg / \HWHM$ from Eq.~(\ref{prop}). In this case, the best relative precision at the maximum oscillation frequency can be derived from Eq.~(\ref{conditiondeux}):
\begin{equation}\label{precision}
   {\delta \dnu \over \dnu}
   \simeq {0.6 \over \amp} {\dnu  \over \larg }
   \simeq {2.4\over  \ampmax}\  \left({\larg \over  \dnu}\right)^{-2} .
\end{equation}
The scaling to the large separation insures a uniform precision throughout the HR diagram. With $\larg\simeq 2\, \dnu$, a 1\% precision on the determination of $\deltanunu$ requires $\ampmax \ge 60$, which is achieved by most of the CoRoT targets (Table~\ref{tablehbr}). Figures \ref{autodeltanunu49933} and \ref{autodeltanunu181420} show the precision we can obtain on the function $\deltanunu$.

\begin{table}
\caption{Degree identification}\label{unzero}
\begin{tabular}{lrcc}
\hline
Star & $\Deltan$ & $\identideg$ & p-value (in\%)\\
\hline
HD49385   &  7  &  2.7 & 0.34\\
HD49933 IR&  8  &  2.0& 2.22\\
HD49933 LR& 10  &  6.0& 0.00\\
HD181420  &  6  &  3.1& 0.10\\
HD181906  &  4  &  0.1& 46.6\\
\hline
\end{tabular}

IR and LR represent respectively the initial and long runs observation on HD 49933, from \cite{2008A&A...488..705A} and \cite{benomar2009}.

\end{table}

\subsection{Disentangling the degree}

Examining the half-separations $\Delta_{01}=\nu_{n,1}-\nu_{n,0}$ and $\Delta_{10}=\nu_{n+1,0}-\nu_{n,1}$, as proposed by R09, requires $\larg$ narrower than $\dnu$.
We have found that $\larg / \dnu=0.75$ provides the best compromise: it is narrow enough to select only 1 pair of modes with degree 0 and 1, and large enough to give an accurate signal-to-ratio.
Setting $\larg / \dnu=0.75$, a 1\% precision on the determination of $\dnu$ requires $\ampmax \ge 430$, which is achieved only for HD49933 and HD49385.

\begin{figure}
\centering
\includegraphics[width=8.5cm]{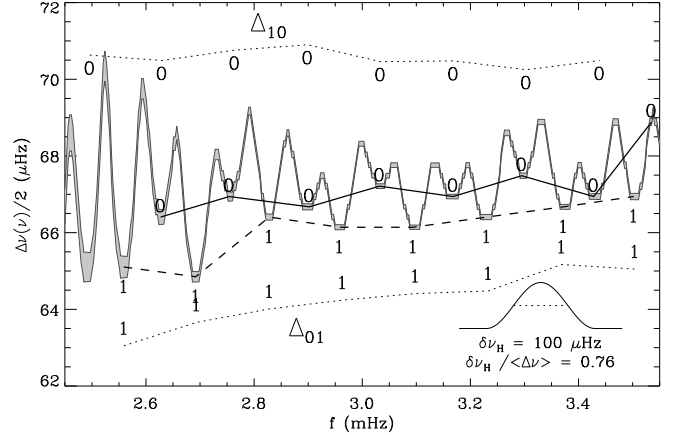}
\caption{Function $\deltanunu$ of the Sun, with a filter width equal to 0.75 times the mean large separation. The extrema of $\deltanunu/2$ do not correspond to the curves $\Delta_{01}$ and $\Delta_{10}$, plotted as dotted curves. The symbols 0 and 1 indicate the location of the eigenfrequencies on the $\Delta_{01}$ and $\Delta_{10}$ curves, and indicate also the corresponding local minima of $\deltanunu$.
\label{autodeltanusoleil}}
\end{figure}

\begin{figure}
\centering
\includegraphics[width=8.5cm]{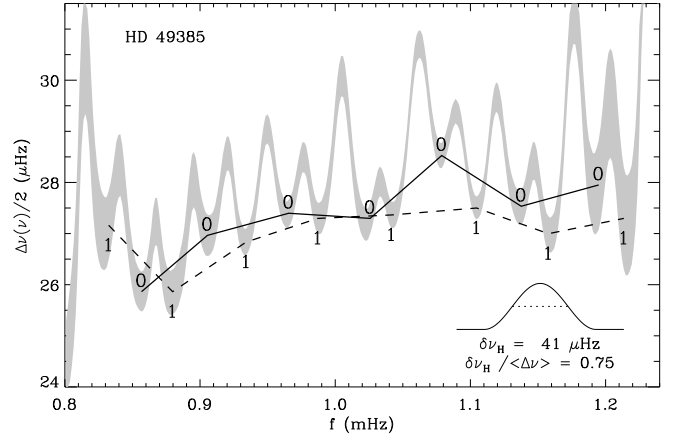}
\caption{Function $\deltanunu$ of HD 49385, with a filter width equal to 0.75 times the mean large separation.
The region in grey encompasses the values $\pm$ the 1-$\sigma$ error bar.
The local minima of $\deltanunu$, here marked by the degree, correspond to the eigenfrequencies mentioned by Deheuvels et al. (2009).
\label{autodeltanano49385}}
\end{figure}

As reported by R09, the different half-large separations are clearly distinguished.
However,  values are correlated within the filter, and mixed with other separations including $\degre=2$ modes. Therefore, we do not consider that the autocorrelation is able to provide a precise measurement of the half-separations. For instance, we cannot reproduce the Solar values (Fig.~\ref{autodeltanusoleil}). However, we clearly show that the local minima  match the eigenfrequencies with high precision. We observed, in the unambiguous cases provided by the Sun, HD49385 (Fig.~\ref{autodeltanano49385}) and models, that the local minima associated with $\degre=1$ are lower than the ones with $\degre=0$. This is due to the fact that, under the assumption of a Tassoul-like spectrum, a narrow filter centered on the $\degre=1$ mode tests the separations $\Delta_{01}$ and $\Delta_{12}=\nu_{n,2}-\nu_{n,1}$, whereas a narrow filter centered on the $\degre=0$ mode tests  the separations $\Delta_{10}$ and $\Delta_{01}$. In the asymptotic formalism, $\Delta_{12}$ is significantly smaller than $\Delta_{10}$ (by an amount of $4\, D_0$).

\begin{figure}
\centering
\includegraphics[width=8.5cm]{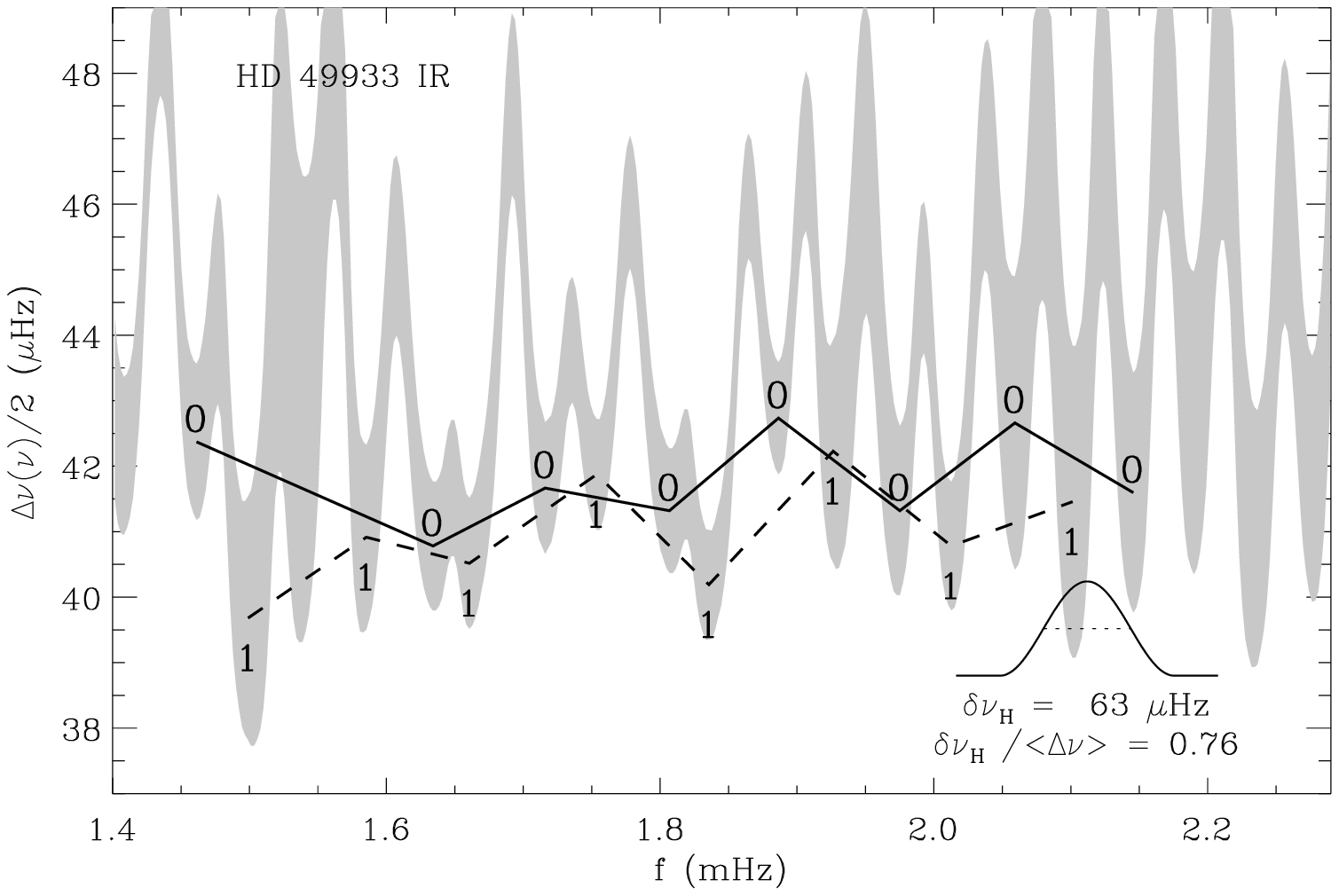}
\includegraphics[width=8.5cm]{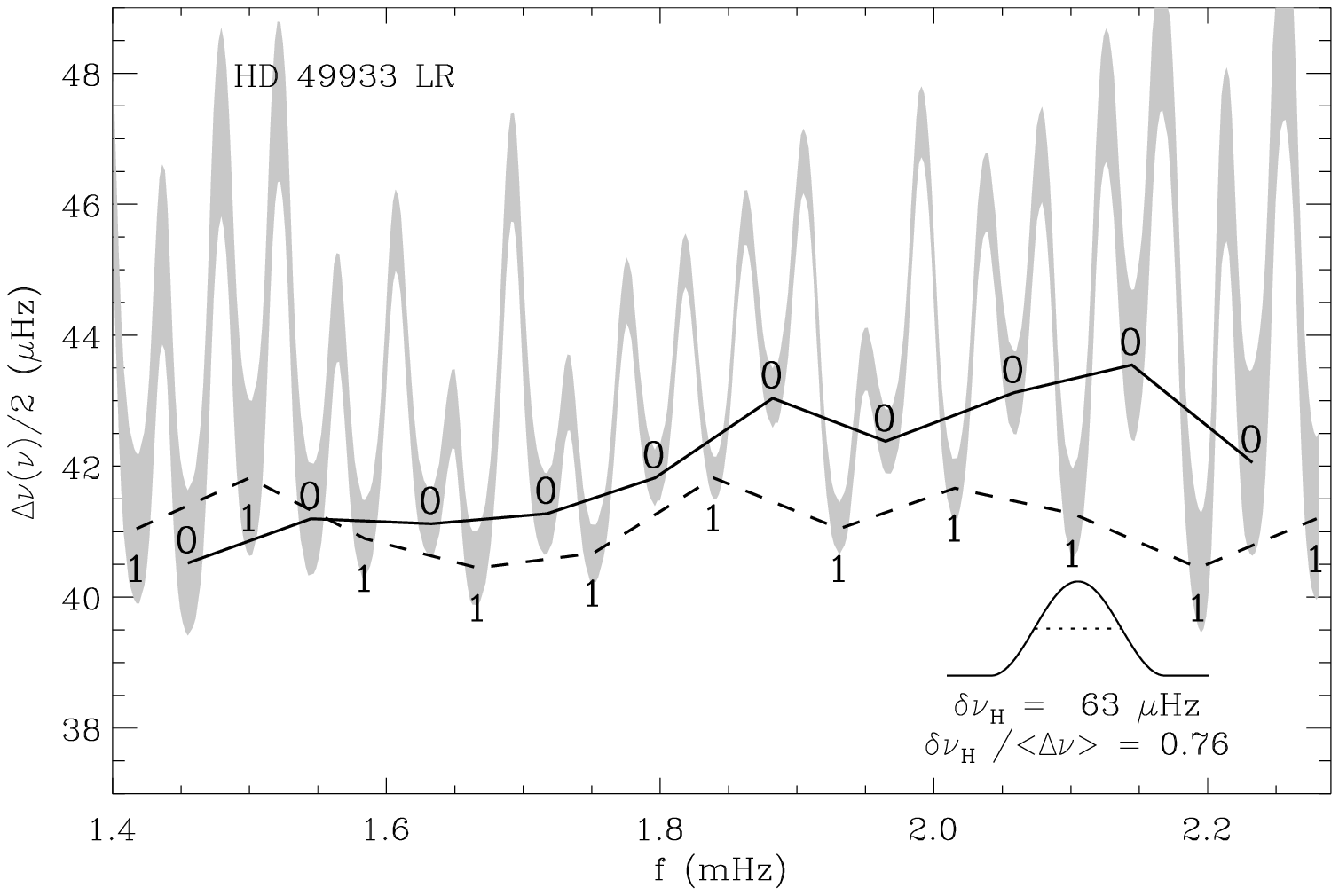}
\caption{Function $\deltanunu$ of HD49933, with a filter width equal to 0.75 times the mean large separation.
Radial and $\degre=1$ modes are identified. The two plots correspond to the 2 runs: the initial run IR lasted 60.7 days and the long run LR 136.9 days.
\label{autodeltanano49933}}
\end{figure}

When the filter is not centered on an eigenmode, it mainly tests the separation $\Delta_{01}$ or $\Delta_{10}$. Therefore, we show that a narrow frequency windowed autocorrelation allows us to distinguish $\degre=0$ from $\degre=1$, which is a crucial issue since many observations  have shown how difficult it can be to distinguish them (\cite{barban2009}, \cite{garcia2009}). The test applied to the first initial run on HD 49933 (Fig.~\ref{autodeltanano49933}) shows that the former mode identification of \cite{2008A&A...488..705A} cannot be confirmed, as also shown by \cite{benomar2009} who analyze a second longer run.

A clear identification requires a signal-to-noise ratio high enough. Again, Eq.~(\ref{precision}) allows us to estimate the autocorrelation amplitude required. In order to distinguish the small separation $D_0$, and considering as a rough estimate that in the mean case $D_0$ represents about 2\% of the large separation, a reliable determination  based on a narrow filter $\larg = \narrow\, \dnu$ requires a maximum amplitude greater than about \ampdeg.

Table~\ref{unzero} summarizes the mean value of the difference $\deltadeg$ between the local minima compared to the 1-$\sigma$ uncertainty $\delta \dnu /2$ of the narrow frequency windowed autocorrelation function (Eq.~\ref{precision}):
\begin{equation}\label{identi}
   \identideg = {1\over \sqrt{\Deltan}} \ \sum_{i=1}^{\Deltan}\ {{\deltadeg}_i \over \delta {\dnu}_i}
\end{equation}
with  $\Deltan$ the number of pairs of modes above the threshold level $\amp = 8$.
Table~\ref{unzero} also provides the probability of obtaining a result as extreme as the observation assuming that the null hypothesis is true.

\begin{figure}
\centering
\includegraphics[width=8.5cm]{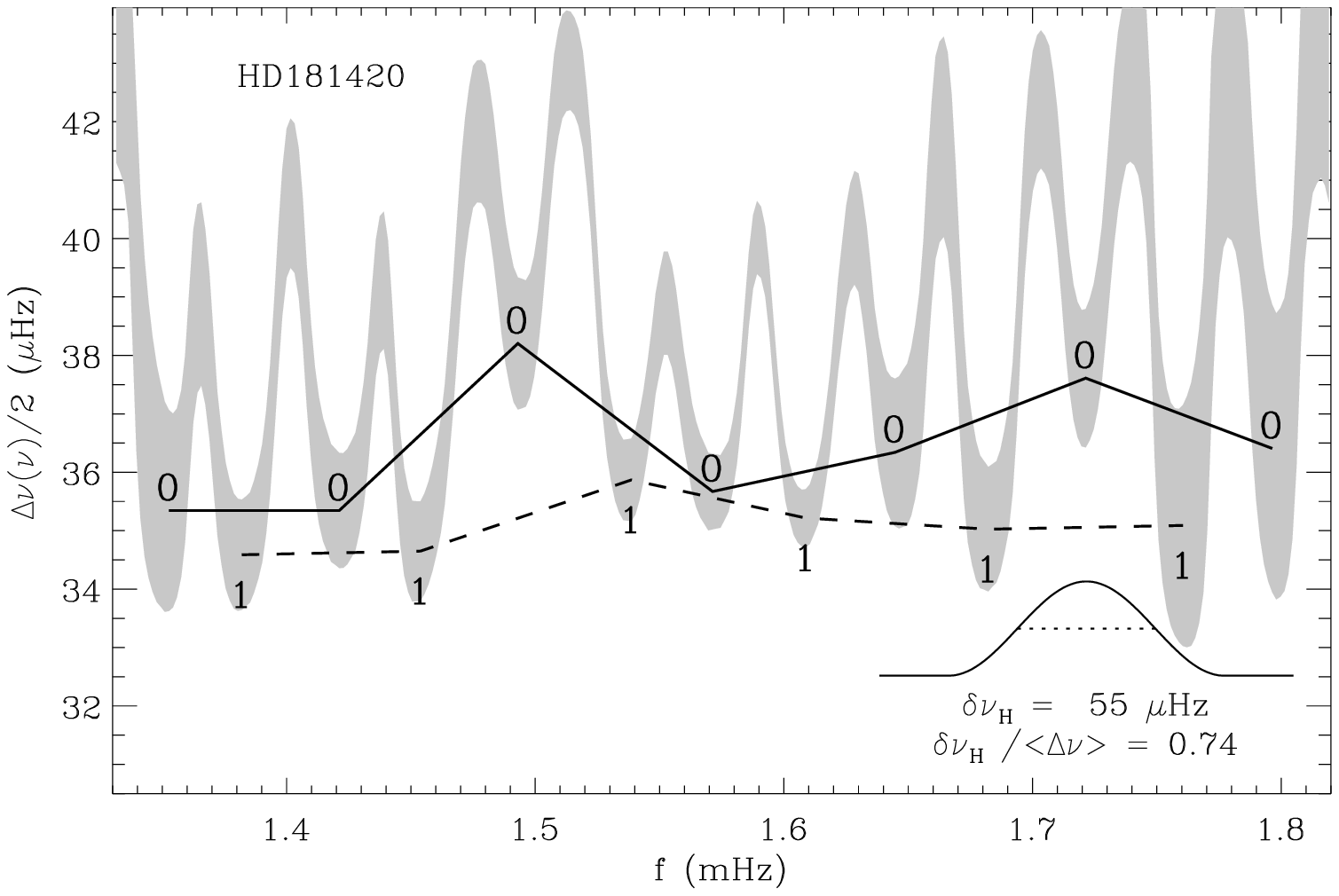}
\caption{Same as Fig.~\ref{autodeltanano49933}, but for HD181420.
\label{autodeltanano181420}}
\end{figure}

\begin{figure}
\centering
\includegraphics[width=8.5cm]{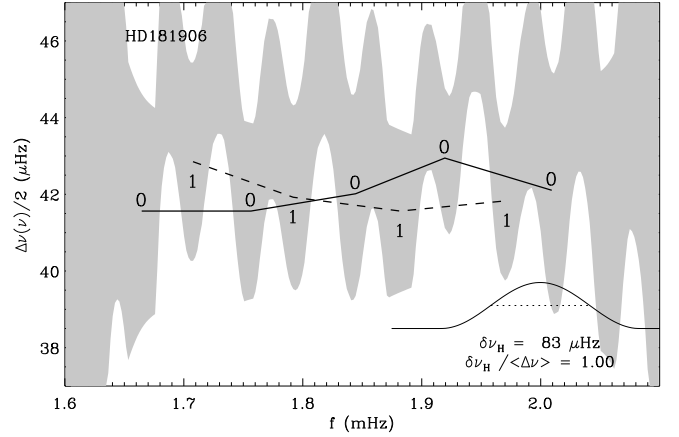}
\caption{Same as Fig.~\ref{autodeltanano49933}, but for HD181906 and with a broader filter. The large uncertainty indicated by the broad grey region shows that the identification for that star is not reliable.
\label{autodeltanano181906}}
\end{figure}

This criterion helps to explain why the ridges can be unambiguously identified in HD 49385 (\cite{deheuvels2009}) and why scenario 1 for HD 181420 must  be preferred (\cite{barban2009}): figures \ref{autodeltanano49933} and \ref{autodeltanano181420} show that the mean difference between the local minimal corresponding to $\degre=0$ or 1 is greater than the error bar of $\deltanunu$. On the other hand, no answer can be given for HD 181906 (\cite{garcia2009}): the low value of $\ampmax$ hampers the calculation of $\deltanunu$ with a narrow filter (Fig.~\ref{autodeltanano181906}).
The limited $\ampmax\simeq 250$ for the initial run on HD49933 and the low value $\identideg$ help to explain the difficulties encountered with the mode identification given in \cite{2008A&A...488..705A}.

\subsection{Ultimate precision on $\deltanunu$}

Obtaining the best time resolution in the \fonction, namely the time resolution $\sampling$ of the time series, requires $ \amp \ge {1.21 / \larg\; \sampling}$ for a filter width $\larg$ (cf. Eq.~\ref{conditionunun}). This relation imposes a strong constraint on the maximum amplitude. Furthermore, in order to investigate the variation $\deltanunu$, the condition has to be satisfied in a frequency range as large as $2\, \HWHM$ around $\centre$. This yields the condition:
\begin{equation}\label{fullresolution}
\amp \ge {4.9 \over \larg\; \sampling} .
\end{equation}
According to this, high precision is easier to reach for high $\larg$ values, hence for low mass stars with a higher large separation. The decrease of the limit with increasing sampling $\sampling$ corresponds to a correlated decrease in resolution.

\subsection{Small separation}

\cite{2006MNRAS.369.1491R} proposed to make use of the autocorrelation function to obtain an independent estimate of the small separation. This method is based on the comparison of the peak amplitude of even or odd orders in the autocorrelation function. With $\amp_n$ the amplitude of the peak of order $n$, it consists of comparing the decreasing  $\amp_{2n}$ values to the increasing $\amp_{2n+1}$ (see Fig.~4 of RV06). Note that $\ampdnu$ corresponds to $\amp_2$. Equality of the interpolated curves $\amp_{2n}$ and $\amp_{2n+1}$ occurs for $n$ of 3 or 4.  Tests made on the available CoRoT data show that values of $\amp_{2n}$ larger than 15 are necessary to apply the method, therefore requiring very large values of $\ampdnu$, larger than about \amppetit.
In the solar case, with the simulation including additional photon noise, the detection limit also occurs at $\HBR \simeq\amppetit$.

\begin{table}
\caption{Threshold levels}\label{relevance}
\begin{tabular}{rp{6.99cm}}
\hline
$\ampmax$  &  detection\\
\hline
$<5.7$ & no detection (10\,\% rejection level)  \\
$<8.0$ & no detection (1\,\% rejection level)  \\
\hline
$10$   & measurement of $\dnumoy$ with 0.5\,\% precision\\
$\ge \ampfit$ & fitting of the modes has proven to be possible\\
$\ge \ampdeg$ & identifying the mode degree has proven to be possible\\
$\ge\amppetit$  & possible estimate of the small separation as in
\cite{2006MNRAS.369.1491R} \\
\hline
\end{tabular}

The threshold values are given for a typical F dwarf. They can be made precise for a given star, according to its specific parameters $\alpha$ and $\gamma$. Levels are lower for red giants (to be estimated with $\gamma \simeq 4$ instead of 10).
\end{table}

\subsection{Threshold levels}

Table \ref{relevance} summarizes the threshold levels for the determination of the seismic parameters with the \fonction.  At low signal-to-noise ratios, namely an $\ampmax$ value lower than 8, the method cannot operate, according to the H$_{0}$ test. Then, the domain where the autocorrelation is highly performing is for $\ampmax$ ranging from 8 (detection limit) to $\simeq\ampfit$, when precise mode fitting becomes possible (HD181906, \cite{garcia2009}). With $\ampmax$ value up to $\ampdeg$, the \fonction\ may be useful for identifying the degree of the modes, under the condition that the oscillation spectrum is close to a Tassoul-like pattern. Larger  $\ampmax$ values allow a more detailed analysis with classical methods such as mode fitting (\cite{2006ESASP1306..377A}).

\begin{figure}
\centering
\includegraphics[width=8.5cm]{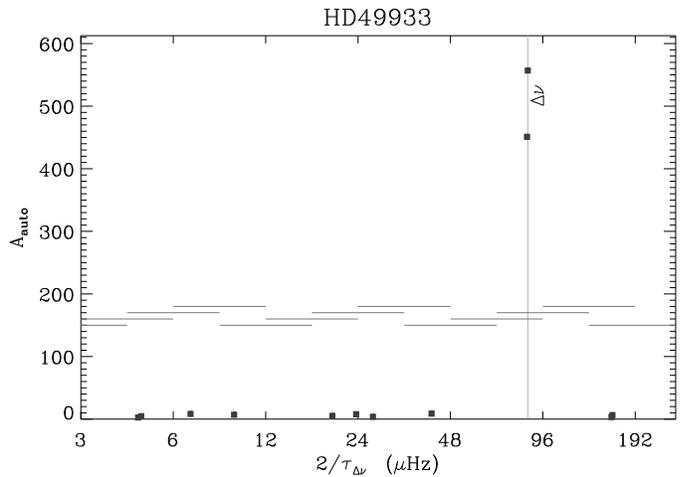}
\caption{Automatic search for the signature of a large separation for HD49933. The grey line indicates the location of the maximum peak, corresponding to the mean large separation of the star. The horizontal segments indicate the ranges corresponding to the 13 initial guess values $\dnuc$.
\label{chercheautodnu49933}}
\end{figure}

\begin{figure}
\centering
\includegraphics[width=8.5cm]{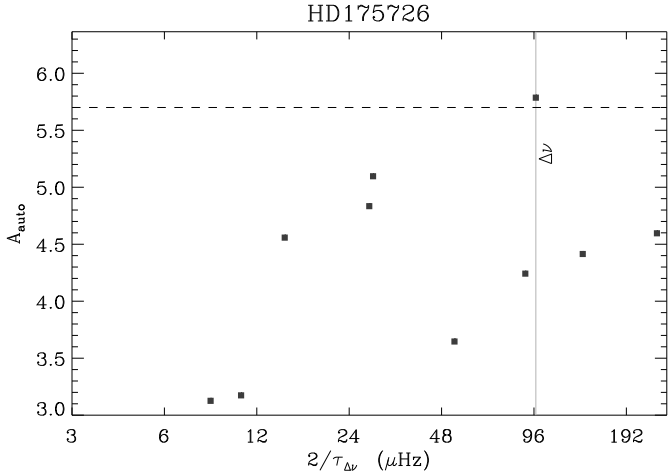}
\caption{Same as Fig.~\ref{chercheautodnu49933}, but for HD175726. The dashed line indicates the 10\,\% rejection limit.
\label{chercheautodnu175726}}
\end{figure}

\begin{figure}
\centering
\includegraphics[width=8.5cm]{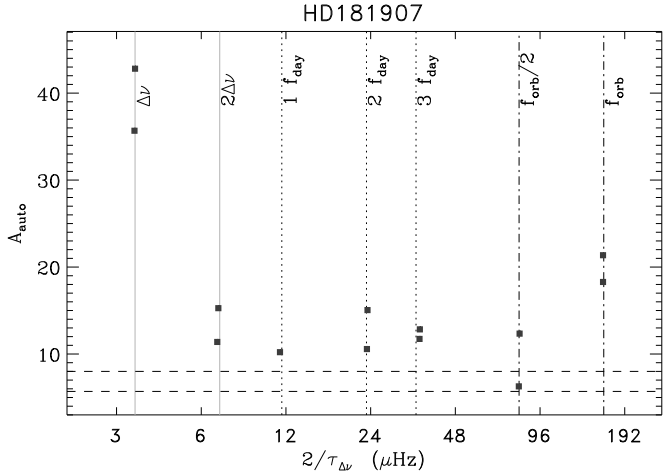}
\caption{Same as Fig.~\ref{chercheautodnu175726}, but for the red giant HD181907. The horizontal dashed lines indicate the 10\,\% and 1\,\% rejection limits. The vertical grey lines  indicate the signature at $\dnu$ and $2\,\dnu$, the dotted line the spurious signatures of the day aliases, and the dot-dashed lines the CoRoT orbital frequency and first subharmonic.
\label{chercheautodnu181907}}
\end{figure}

\begin{table*}
\caption{\Kepler\ performance\label{amplikepler}}
\begin{tabular}{llrrrrrrrrrrrrr}
\hline
        &     &       & &\multicolumn{5}{c}{90-day run} & & \multicolumn{5}{c}{4-year run}\\
stellar  &type & $B-V$ & &\multicolumn{5}{c}{$m\ind{V}$ magnitude}& &\multicolumn{5}{c}{$m\ind{V}$ magnitude}\\
type as  &     &       & &  9  &    10 &    11 & 12    &   13 & &    9  &    10 &    11 &    12 & 1   3\\
\hline
HD181907 &G8III&  1.09 & &  23 &    23 &    23 &    23 &    22& &   379 &   378 &   377 &   373 &   364\\
\hline
HD 49385 & G0IV&  0.51 & & 167 &    53 &    14 &   3.6 &   0.9& &  2709 &   864 &   233 &    58 &    14\\
\hline
HD 49933 & F5V &  0.35 & &  44 &    11 &   2.9 &   0.7 &   0.2& &   715 &   188 &    46 &    11 &   2.7\\
HD181420 & F2V &  0.40 & &  36 &    10 &   2.5 &   0.6 &   0.1& &   598 &   161 &    40 &   9.8 &   2.3\\
HD181906 & F8V &  0.43 & &  27 &   6.9 &   1.7 &   0.4 &   0.1& &   446 &   111 &    27 &   6.5 &   1.5\\
HD175726 & G0V &  0.53 & & 3.3 &   0.8 &   0.2 &   0.0 &   0.0& &    54 &    13 &   3.3 &   0.8 &   0.2\\
\hline
\end{tabular}

Maximum amplitude $\ampmax$ for \Kepler\ performance on CoRoT-like targets with varying magnitudes.
\end{table*}

\section{Discussion\label{discussion}}

\subsection{Automated determination of the large separation}

Autocorrelation may provide an effective automatic determination of the large separation when nothing is known about the star, as can be the case for a \Kepler\ target. As shown previously, testing the autocorrelation around $\Tdnu$ requires a cosine filter $\larg = \alpha\gamma\; \dnu \simeq 10.5\; \dnu$, near the frequency $\numax$. In order to perform the test in fully blind conditions, in a frequency range simultaneously encompassing giant and dwarf stars, $\numax$ is derived from the scaling law given by Eq.~(\ref{scalingnumax}).

The automatic test consists of analyzing the autocorrelation of the time series for a set of time shifts $\Tdnuc$ in geometrical progression. We performed the automatic autocorrelation test with 13 values of $\dnuc=2/\Tdnuc$, varying from 3 to 192\muHz\ with a geometric ratio $\coefauto$ equal to $\sqrt{2}$; $\dnuc$ can be considered as an initial guess of the large separation. For each initial value $\dnuc$, we explored the range $[\dnuc/ \coefauto, \dnuc \, \coefauto]$ of the autocorrelation for
3 frequency ranges of the Fourier spectrum centered respectively around $\numax$ and $\numax\pm\larg/2$. We finally derived the large separation from the maximum amplitude $\ampauto$ calculated for each $\dnuc$ initial guess. Comparison of the different $\ampauto$ is made possible by the scaling provided by Eq.~(\ref{normalisation}). Figure~\ref{chercheautodnu49933} shows the result for HD49933.

We also tested the automatic test with the stars with the lowest $\ampmax$, namely HD175726 and HD181907.
For HD175726, the single value exceeding the 10\,\% rejection level occurs at 97\muHz\ (Fig.~\ref{chercheautodnu175726}). This value of $\dnu$ agrees with the solution proposed by Mosser et al. (2009). This detection is poor since a significance level of 10\,\% means that the posterior probability of the null hypothesis is at least 38\,\% according to \cite{{2009arXiv0906.0864A}}. However, the automatic detection can be refined with a dedicated search with a more precise grid of analysis. In the case of HD175726, the clear identification of an excess power centered at 2\,mHz first allows us to better estimate the parameters for searching $\dnu$ and, second, gives a further indication that the measurement is reliable thanks to Eq.~(\ref{scalingnumax}).

The amplitude $\ampauto$ of the automatic test is found to be close to the maximum amplitude $\ampmax$. Only limited fine tuning around the automatically fixed parameters is needed to optimize the result.  Mosser et al. (2009) have mentioned the difficulty of determining the large separation with other methods. The autocorrelation method proves to be powerful for a rapid estimate of the large separation; rapidity means here a few seconds of CPU time for the Fourier spectrum of the CoRoT time series, followed by a few seconds of CPU time for the automatic search with the autocorrelation, with a common laptop.

We verified that the method is effective for all other solar-like targets: it gives one single answer for $\dnu$, and does not deliver any false positives. The case of red giants requires a dedicated analysis.

\begin{figure}
\centering
\includegraphics[width=8.5cm]{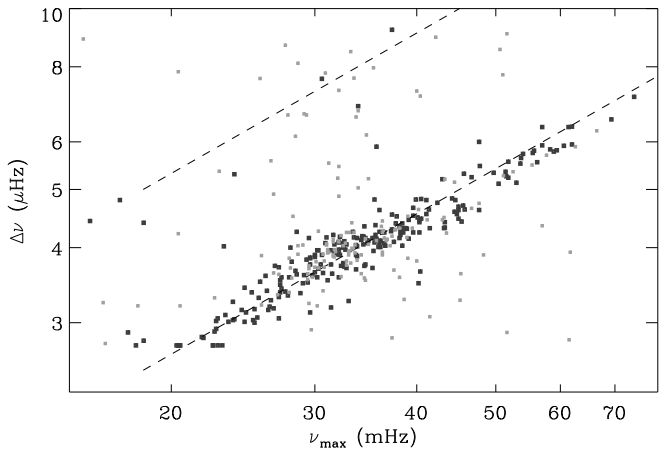}
\caption{Large separation, automatically measured for a set of 392 red giants analyzed in \cite{hekker2009}, as a function of the maximum oscillation frequency $\numax$. Large black squares indicate positive detection; small grey squares correspond to unreliable cases. The dashed line indicate the global fit described by Eq.~\ref{exposantridge}; a few cases correspond to the automatic detection of twice the large separation.
\label{geantes}}
\end{figure}

\subsection{Red giants}

We tested the method on the CoRoT red giant target HD 181907 (\cite{carrier2009}).
With a large  $\HBR$ for this target (about 2) and a large $\ampmax$ autocorrelation signal, the large separation is easily found around 3.5\,\muHz, but the detection is polluted by many values clearly above the 1\,\% rejection limit (Fig.~\ref{chercheautodnu181907}). All these spurious detections are caused by artefacts: detection of the double of the large separation; detection of the diurnal frequency and its harmonics; detection of the CoRoT orbital frequency and half its value. We checked that the detections at high harmonics of the diurnal frequency are due to residuals of the window function (\cite{mosser2009}); they are introduced by the link between $\numax$ and $\dnu$ indicated by Eq.~\ref{scalingnumax}.

For HD 181907, Eq.~(\ref{propmax}) is valid within 30\,\%. As discussed in Section~\ref{analysis}, the discrepancy compared to solar-like stars is due to the fact that the mode envelope of red giants is narrower than in solar-like stars.

The automated determination of the large frequency has been also tested on a set of 392 giants observed in the CoRoT field dedicated to exoplanetary science and analyzed by \cite{hekker2009}. The method proves to be efficient and rapid. It provides a clear advantage since it gives a quantified reliability thanks to the use of the H$_{0}$ test. We present in Figure~\ref{geantes} the results obtained for these giants. The maximum oscillation frequency was calculated by \cite{hekker2009}. The amplitude of the autocorrelation signal allows us to clearly discriminate artefacts from reliable detection (60\,\% of the targets). The relative precision in the mean large separation is much better than 1\,\%, according to Eq.~\ref{conditionmoyenne}. After correction of the stars for which the double of the large separation is preferably automatically detected, we define from the fit of the relation between the large separation and the location of the maximum signal a power law varying as:
\begin{equation}\label{exposantridge}
\dnu \simeq (0.26 \pm 0.015)\  \numax^{0.78\pm 0.03} \ \hbox{(frequencies in $\mu$Hz)}.
\end{equation}
This law for giants is in agreement with Eq.~(\ref{scalingnumax}) based on dwarfs and with \cite{hekker2009}.

\subsection{\Kepler\ data}

The \Kepler\ mission compared to CoRoT will provide different photometric performance, on dimmer targets but in some cases with longer observation duration (\cite{2007CoAst.150..350C}). According to \Kepler\ performance (\cite{KASC0008_5}), the noise level is about 0.92, 10.2 and 144\,ppm$^2\, \mu$Hz$^{-1}$ for targets of V magnitude respectively equal to 9, 11.5 and 14.

We can extrapolate the performance obtained with \Kepler\ on targets similar to the ones observed by CoRoT, but of magnitude 9 to 14, after a 4-year long observation. Table~\ref{amplikepler} gives the amplitude $\ampmax$ for targets observed during typical 90-day or 4-year long runs. According to the expected performance in 90-day runs, the brightest F-type or the class IV targets will have a signal-to-noise ratio high enough to derive information on the large separation.  In a 4-year run, the brightest G dwarfs will deliver a clean seismic signature. On the other hand, faint F targets will have fully exploitable Fourier spectra that will require a precise mode fitting for the most complete seismic analysis. The performance for giants appears to be almost independent of the magnitude, since the  contribution of photon noise is negligible at low frequency.

We can compare this approach to the hare-and-hounds exercises performed by \cite{2008AN....329..549C}. The asteroseismic goal of \Kepler\ is principally to derive information on stars hosting a planet, by the determination of the large separation. Compared to global fitting, the autocorrelation function gives a more rapid and direct answer.

The autocorrelation benefits from the rapid cadence (32 s) provided by CoRoT in the seismology field. \Kepler\ will provide 2 cadences, at 1 or 30\,min. This yields a lower resolution in time, hence a lower precision on the expected results.

\section{Conclusion\label{conclusion}}

\cite{2006MNRAS.369.1491R} have proposed a method for estimating large and small separations from the analysis of the autocorrelation function. \cite{roxburgh2009} has extended the method to determine the variation of the large separation. In this paper, we have developed and quantified the method, relating the amplitude of the correlation peak at time shift $\Tdnu=2/\dnu$ to various parameters.

We have scaled the autocorrelation to the white noise contribution, so that we were able to relate the autocorrelation signal to the mean seismic height-to-background ratio  $\HBR$ that measures the relative power density of the signal compared to noise and to background signals. This  empirical relation is precise to about 15\% for solar-like stars. $\HBR$ aggregates the influence of unknown parameters such as the mode lifetimes, the star inclination (that governs the modes visibility) or the rotational splitting. On the other hand, all these unknown parameters complicate and slow down  the fitting of individual eigenfrequencies. Therefore, the \fonction\ shows here a possible advantage in terms of speed.

The \fonction\ gives a direct measurement of the mean large separation. Compared to other methods, the estimate is accurate and simple, with an intrinsic threshold value, with error bars, and without any modeling of the other components of the Fourier spectrum (granulation or activity). Furthermore, when the signal-to-noise ratio is high enough, the \fonction\ allows the measurement of the variation of the large separation with frequency, without any mode fitting. This is a key point for stellar radius measurement.
Previous works have shown the difficulty to disentangle $\degre=0$ from $\degre=1$ modes in oscillation spectra of F stars observed in photometry. We have verified that, for high signal-to-noise ratio Fourier spectra, the autocorrelation analysis can provide an unambiguous identification of the mode degree for a solar-like oscillation spectrum.

We have defined a method for the automatic determination of the large separation, which is efficient at low signal-to-noise ratio, even if no information is known for the star.
We have determined that the width of the cosine filter used in the method that optimizes the \fonction\ is very close to full-width at half-maximum of the mode envelope (ratio about 1.05). We have also checked that the performance of the method increases linearly with the duration of the time series.
With very limited CPU time (a few seconds), this method delivers the mean large separation of a target. It requires no information on the star; it just relies on the assumption that the location of the excess power and its width are related to the large separation by a scaling law, what is verified for red giants and solar-like stars. Finally, we were able to investigate in a simple manner the capability of \Kepler.

We are confident that the autocorrelation method will be of great help in analyzing high duty cycle time series as a complement to the Fourier analysis. As noticed by \cite{1999A&A...343..608F}, the autocorrelation signal gives a clear signature since the autocorrelation delay, namely four times the stellar acoustic radius (about 4 to 8 hours for an F dwarf), is much shorter than the mode lifetime (a few days). This allows each wavepacket to properly correlate with itself after a double travel along the stellar diameter, so that the autocorrelation integrates phased responses over the total duration time. In the Fourier spectrum, on the contrary, interference between the short-lived wavepackets observed in the time series produce a complicated pattern. But Fourier analysis still remains required for the precise determination of the eigenfrequencies derived from an accurate mode fitting.

\appendix\section{Performance of the autocorrelation}

\subsection{Square module of the autocorrelation}

\begin{figure}
\centering
\includegraphics[width=8.5cm]{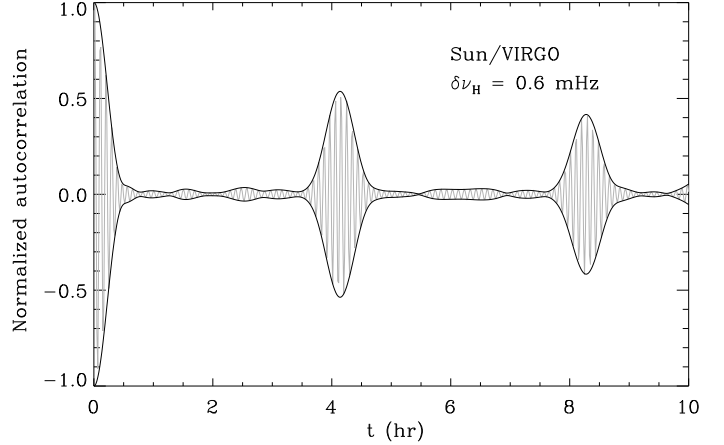}
\caption{Comparison of $\pm\left|\correl\right|$ (black curves) and $\correlc$ (grey line), both normalized to 1 at time shift 0,  calculated for the narrow frequency windowed solar Virgo spectrum.
\label{correl_env}}
\end{figure}

The \fonction\ presented in Section~\ref{autocorrelation} is defined to directly give the envelope of the autocorrelation. Since negative frequencies are omitted, the \fonction\ is related to the canonical autocorrelation $\correlc$, that includes positive and negative frequencies of the Fourier spectrum,  by:
\begin{equation}\label{defautocor}
\amp \propto \left| \correl \right|^2 \hbox{ \ and \ }
\left| \correl \right|  = \sqrt{\correlc^2 + \hilbert ( \correlc ){}^2}
\end{equation}
where $\hilbert$ is the Hilbert transform. Fig.~\ref{correl_env} shows the difference between $\correl$ and $\correlc$, both functions include the contribution of a narrow Hanning filter.

\subsection{Autocorrelation peak}

\begin{figure}
\centering
\includegraphics[width=8.5cm]{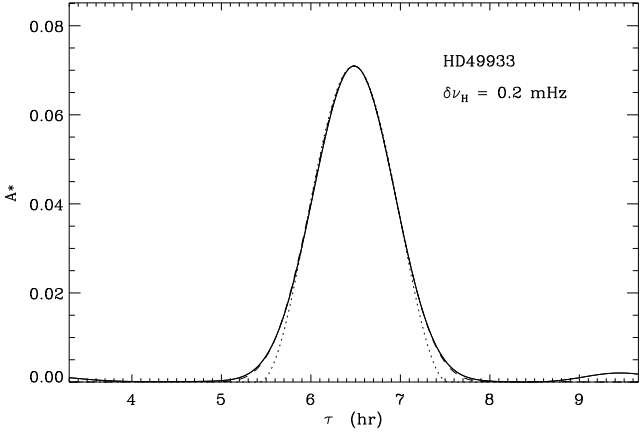}
\caption{Autocorrelation peak ${\ampdnu}^\star$ (solid line) normalized to 1 at time shift 0, theoretical profile (dashed line) and cosine model (dotted line) calculated from the HD 49933 power spectrum and a filter width $\larg = 0.2$\,mHz.
\label{autoforme}}
\end{figure}

The shape of the autocorrelation peaks is given by the Fourier transform of the Hanning filter, which can be expressed as the sum of 3 components (\cite{maxlac}):
\begin{equation}\label{peakhanning}
\mathcal{H} (\tau) \propto
2\,  \sinc (\larg \tau)
+ \sinc(\larg \tau+1)
+ \sinc(\larg \tau-1)
\end{equation}
with $\sinc X = \sin (\pi X) /\pi X$.
This global shape may be more simply modeled with a single cosine shape (Fig.~\ref{autoforme}). In order to enhance the precision of the fit in the upper part of the autocorrelation peak, the full width at half maximum $\largpic$ of this cosine fit is:
\begin{equation}\label{demolargeur}
 \largpic = {\beta \over \larg}  \mathrm{ \ with \ } \beta \simeq 0.763.
\end{equation}
The flanks of the peak are not well fitted, which is unimportant compared to the fact that the fit above half-maximum performs well.  $\largpic$ is much greater than the resolution time $\sampling$.

Precise determination of the large separation requires precise location of the peak maximum. In order to estimate the performance, we describe a peak as:
\begin{equation}
\left\lbrace
\begin{array}{lll}
 S(t)  = \displaystyle{A\over 2} \ \left[1+ \cos \pi \displaystyle{t-\Tdnu\over \largpic} \right]
 & \mathrm{\ for } & |t| \le \largpic, \\
 S(t) = 0 & \mathrm{\ for } & |t| > \largpic.
\end{array}
\right.
\end{equation}
This fit shows variation:
\begin{equation}
   \diff S\ind{A}   = - \displaystyle{\pi A\over 2 \largpic} \ \sin \pi \displaystyle{t-\Tdnu\over \largpic} \ \diff t .
\end{equation}
We can compare the variation of the signal peaking at amplitude $\amp$ to the maximum variation of a noise contribution of amplitude $b$. At a time shift $\Delta t$ from the maximum, the signal variation and the maximum noise contribution are:
\begin{equation}
\left\lbrace
\begin{array}{rcc}
 \Delta S\ind{A}  &= \displaystyle{\amp\over 2} & \left(\pi \displaystyle{\Delta t\over \largpic} \right)^2 , \\
 \Delta S\ind{b}  &= \displaystyle{b \over 2} &   \displaystyle{\Delta t \over \largpic} . \\
\end{array}
\right.
\end{equation}
The precise identification of the signal maximum requires $ \Delta S\ind{A} \ge \Delta S\ind{b}$, which translates into the condition:
\begin{equation}\label{condition}
\pi\; \amp\; \Delta t \ge  b\; \largpic .
\end{equation}
It is possible to interpret this condition as follows.

\subsection{Precision on the mean large separation}

If the amplitude is not large enough, then Eq.~\ref{condition} defines a resolution $\Delta t$, hence a limited precision on the large separation:
\begin{equation}\label{conditiondeux}
   {\delta \dnu \over \dnu} = {\Delta t\over \Tdnu} =  {\beta \over 2\pi}\; {b\over  \amp} \; {\dnu \over  \larg}  .
\end{equation}
We can set, as a limit to detection, $b =5$ and $\amp = 8$.

\subsection{Full resolution for the measurement of $\deltanunu$}

In order to recover the full time resolution $(\Delta t = \sampling)$, the amplitude must satisfy $\amp \ge \ampdt$, with the definition:
\begin{equation}\label{conditionun}
   \ampdt  = {b \over \pi}{\largpic \over \sampling} = {b \beta \over \pi}{1 \over \larg\, \sampling}  .
\end{equation}
With $b=5$:
\begin{equation}\label{conditionunun}
   \ampdt  \simeq  {1.21 \over \larg\, \sampling} .
\end{equation}

\begin{acknowledgements}
This work was supported by the Centre National d'Etudes Spatiales (CNES). It is based on observations with CoRoT. Solar data were obtained from SOHO, a mission of international collaboration between ESA and NASA. \\
The work has benefitted from simulating discussions among asteroseismologists working on the excellent CoRoT data. BM thanks Ian Roxburgh for motivating discussions, and John Leibacher for helpful comments. We thank Saskia Hekker and Caroline Barban for providing us with the red giant data and the values of the maximum frequency plotted in Fig.~\ref{geantes}.

\end{acknowledgements}

\end{document}